\documentclass[12pt, epsfig]{article}
\pdfoutput=1
\usepackage{amsmath,amsfonts,amssymb,amsthm,amstext,amscd,eucal,graphicx,enumitem}
\usepackage[all]{xy}
\usepackage{dsfont}
\usepackage{hyperref}
\usepackage{amsmath}
\usepackage{slashed}
\usepackage{color}
\usepackage{epsfig}%
\usepackage[active]{srcltx}
\usepackage{enumitem}
\setlist{nolistsep}

\newcommand{\dd}{{\rm d}}

\makeatletter \@addtoreset{equation}{section}

\makeatletter\renewcommand\section{\@startsection {section}{1}{\z@}%
                                   {-3.5ex \@plus -1ex \@minus -.2ex}
                                   {2.3ex \@plus.2ex}%
                                   {\normalfont\large\bfseries}}
\renewcommand\subsection{\@startsection{subsection}{2}{\z@}%
                                     {-3.25ex\@plus -1ex \@minus -.2ex}%
                                     {1.5ex \@plus .2ex}%
                                     {\normalfont\bfseries}}

\newcommand{\be}{\begin{equation}}
\newcommand{\ee}{\end{equation}}
\newcommand{\bea}{\begin{eqnarray}}
\newcommand{\eea}{\end{eqnarray}}
\newcommand{\bse}{\begin{subequations}}
\newcommand{\ese}{\end{subequations}}
\newcommand{\beqa}{\begin{eqnarray}}
\newcommand{\eeqa}{\end{eqnarray}}
\newcommand{\beqar}{\begin{eqnarray*}}
\newcommand{\eeqar}{\end{eqnarray*}}
\newcommand{\bi}{\begin{itemize}}
\newcommand{\ei}{\end{itemize}}
\newcommand{\bn}{\begin{enumerate}}
\newcommand{\en}{\end{enumerate}}

\newcommand{\ba}{\begin{array}}
\newcommand{\ea}{\end{array}}
\newcommand{\bc}{\begin{center}}
\newcommand{\ec}{\end{center}}
\newcommand{\bal}{\begin{align}}
\newcommand{\eal}{\end{align}}

\newcommand{\nn}{\nonumber}

\def\dd{\textrm{d}}
\def\nn{\nonumber \\}

\newcommand{\vol}{\mbox{vol}}

\definecolor{darkgreen}{rgb}{0,0.3,0}
\definecolor{darkblue}{rgb}{0,0,0.3}
\definecolor{darkred}{rgb}{0.7,0,0}

\begin{document} 

\begin{titlepage}

\vfill
\begin{flushright}
APCTP-Pre2017-006 \\
DMUS--MP--17/06 
\end{flushright}

\vfill

\begin{center}
   \baselineskip=16pt
   {\Large \bf Calibrated Entanglement Entropy}
   \vskip 2cm
     I. Bakhmatov$^{a,b}$, N. S. Deger$^c$, J. Gutowski$^d$, E. \'O Colg\'ain$^{a}$, H. Yavartanoo$^{e}$
       \vskip .6cm
             \begin{small}
             \textit{$^a$ Asia Pacific Center for Theoretical Physics, Postech, Pohang 37673, Korea}\\
                 \vspace{3mm} 
             \textit{$^b$ Institute of Physics, Kazan Federal University, Kremlevskaya 16a, Kazan, 420111, Russia}\\
             	 \vspace{3mm}    
             \textit{$^c$ Department of Mathematics, Bogazici University, Bebek, 34342, Istanbul, Turkey}\\
                 \vspace{3mm} 
             \textit{$^d$ Department of Mathematics, University of Surrey Guildford, GU2 7XH, UK}\\
                 \vspace{3mm}
             \textit{$^e$ State Key Laboratory of Theoretical Physics, Institute of Theoretical Physics, \\ Chinese Academy of Sciences, Beijing 100190, China}
                 \vspace{3mm}
             \end{small}
\end{center}

\vfill \begin{center} \textbf{Abstract}\end{center} \begin{quote}
The Ryu-Takayanagi prescription reduces the problem of calculating entanglement entropy in CFTs to the determination of minimal surfaces in a dual anti-de Sitter geometry. For 3D gravity theories and BTZ black holes, we identify the minimal surfaces as special Lagrangian cycles calibrated by the real part of the holomorphic one-form of a spacelike hypersurface. We show that (generalised) calibrations provide a unified way to determine holographic entanglement entropy from minimal surfaces, which is applicable to warped AdS$_3$ geometries. We briefly discuss generalisations to higher dimensions. 
\end{quote} \vfill

\end{titlepage}



\section{Introduction}
Holographic entanglement entropy, as originally conceived by Ryu-Takayanagi \cite{Ryu:2006bv, Ryu:2006ef} (RT) has been an unmitigated success. For static configurations, it recasts the problem of determining entanglement entropy in 2D CFTs \cite{Holzhey:1994we, Calabrese:2004eu, Calabrese:2009qy} as a calculation of the area of a co-dimension two minimal surface in an AdS$_3$ bulk spacetime. The great appeal of this approach is that it readily generalises to higher-dimensional CFTs and their AdS duals, as well as more generic field theories with gravity duals. Moreover, a covariant generalisation  \cite{Hubeny:2007xt} permits one to start addressing the time-dependence of entanglement entropy. To date, various explanations of the holographic prescription have appeared in the literature \cite{Casini:2011kv, Hartman:2013mia, Faulkner:2013yia, Lewkowycz:2013nqa, Dong:2016hjy, Sheikh-Jabbari:2016znt}, leading to great confidence in the relation. 

In practice geometric calculations, especially in higher dimensions, are still tricky. Restricted to symmetric entangling surfaces, namely balls or strips, the higher-dimensional problem retains some of the simplicity of the 3D problem. However, for generic configurations, the recognised prescription involves solving second-order equations in a bid to identify minimal surfaces. In this paper, we import \textit{calibrations}  \cite{Harvey:1982xk, calibrations} from the mathematics literature to aid the identification.  Calibrations provide a mechanism to determine minimal surfaces in curved space and received early attention in the string theory context of Calabi-Yau compactifications \cite{Becker:1995kb, Becker:1996ay, Gibbons:1998hm} \footnote{See also \cite{Gauntlett:1998vk, Gauntlett:2003cy, Martelli:2003ki, HackettJones:2004yi, Martucci:2005ht, Gauntlett:2006ai, deFelice:2017mhm} for applications of calibrations to supersymmetric branes.}. The connection to supersymmetry is not so surprising as calibrations admit a spinorial construction \cite{calibrations_spinor}, yet may be defined in the absence of supersymmetry. We note that it was recently shown that RT minimal surfaces in 3D are supersymmetric and one can deduce the surfaces without solving the geodesic equation \cite{Colgain:2016lxi}. 

One catch of using calibrations is that they are only defined for Riemannian manifolds and not pseudo-Riemannian counterparts. As a result, we must split our spacetime into a timelike direction and a transverse spacelike hypersurface where one may define a calibrated cycle. Once this is done, the approach hinges on identifying a closed differential form to define the calibration, which even in the presence of flux, may be replaced by a ``generalised   calibration" \cite{Gutowski:1999iu, Gutowski:1999tu} that is no longer closed. This programme is easily implemented in 3D, where the hypersurface is a 2D Riemann surface, which is necessarily a K\"{a}hler manifold. This leaves two natural candidates for calibrations: the K\"{a}hler two-form, or volume form in this case, and the holomorphic one-form, which may be used to define a \textit{special Lagrangian} (sLag) cycle. Since we are looking for a minimal curve in 3D, the latter is the obvious candidate. 

Over the past few years we have witnessed an increased interest in holographic entanglement entropy in the context of spacetimes that are not asymptotically AdS \cite{Li:2010dr, Maldacena:2012xp, Anninos:2013nja, Gentle:2015cfp}. In particular, one of the simplest departures from the norm involves warped AdS$_3$ spacetimes or black holes. In this context, the dual theory is sensitive to the asymptotic boundary conditions and depending on them, the theory may be warped CFT, with a single copy of Virasoro symmetry and a $U(1)$ Kac-Moody algebra \cite{Detournay:2012pc}, or a more usual CFT with two copies of the Virasoro algebra \cite{Anninos:2008fx, Compere:2014bia}. In the literature, one encounters different proposals for the holographic entanglement entropy. Previously, it has been suggested to identify geodesics in warped AdS$_3$ \cite{Anninos:2013nja}, while more recently, the Lewkowycz-Maldacena procedure \cite{Lewkowycz:2013nqa} has been tailored to the case where the dual theory is conjectured to be a CFT \cite{Song:2016pwx, Song:2016gtd}. Regardless of the procedure, provided there is a minimal surface to be determined, we will demonstrate that calibrations do the job. 

Therefore, in this work, where we focus on 3D spacetimes, we put holographic entanglement entropy in both locally AdS$_3$ and warped AdS$_3$ spacetimes on an equal footing. To do so, we eschew solving the geodesic equation and instead identify a spacelike hypersurface, which allows us to identify a sLag cycle. For massless, static and rotating BTZ black holes, we show that the sLag calibration conditions can be directly solved to find the required minimal surfaces. Furthermore, we demonstrate for warped AdS$_3$ black holes, and their dual putative CFTs, that the sLag calibration corresponds to a generalised  calibration, where the calibration is no longer closed, but proportional to the flux sourcing the warping. In contrast, for warped CFTs, it is appropriate to simply consider calibrations. Explicitly, we show the former for warped black hole solutions to a consistent truncation of 10D supergravity \cite{Detournay:2012dz}. 

One can neatly summarise our findings on holographic entanglement entropy $S_{\textrm{EE}}$, as 
\be
\label{SEE}
S_{\textrm{EE}} = \frac{1}{4 G_3} \int_{\textrm{sLag}} \textrm{Re} (\varphi), 
\ee 
where $\varphi = e^{i \chi} \Omega$  is the holomorphic one-form on a spacelike hypersurface with a phase $\chi$ that is fixed appropriately. 

The structure of this short note is as follows. In section \ref{sec:calibrations} we review calibrations, before applying this technology to BTZ black holes \cite{Banados:1992wn, Banados:1992gq} in section \ref{sec:BTZ}. In section \ref{sec:warped}, we demonstrate that the minimal curves for a class of warped AdS$_3$ black holes dual to CFTs correspond to generalised sLag calibrations, while in section \ref{sec:warped_geo}, we identify minimal surfaces using calibrations, before concluding with a discussion of the utility of the method in higher dimensions. In the appendix, we present a solution to the geodesic equation for rotating BTZ.

\section{Review of calibrations}
\label{sec:calibrations}
We begin with a review of calibrations following  \cite{Harvey:1982xk} and its extension to generalised   calibrations \cite{Gutowski:1999iu, Gutowski:1999tu}. We consider a Riemannian manifold $\mathcal{M}$ and a \textit{closed} exterior $p$-form $\varphi$ with the property that 
\be
\label{cal1}
\varphi |_{\xi} \leq \textrm{vol}_{\xi}, 
\ee
for all oriented tangent $p$-planes $\xi$ on $\mathcal{M}$. Then, any compact oriented $p$-dimensional submanifold, or cycle, $\mathcal{N}$ of $\mathcal{M}$ with the property that 
\be
\label{cal2}
\varphi |_{\mathcal{N}} = \textrm{vol}_{\mathcal{N}}, 
\ee
is guaranteed to be a volume minimising submanifold in its homology class, or put more mathematically, $\vol (\mathcal{N}) \leq \vol(\mathcal{N}')$ for any $\mathcal{N}'$ such that the boundaries agree $\partial \mathcal{N} = \partial \mathcal{N}'$ and $[\mathcal{N} - \mathcal{N}'] = 0$ in $H_{p} (\mathcal{M}; \mathbb{R})$. To appreciate this fact, one should simply note that 
\be
\vol(\mathcal{N}) = \int_{\mathcal{N}} \varphi = \int_{\mathcal{N}'} \varphi \leq \vol(\mathcal{N}'), 
\ee
where the first equality and last inequality follow from the above equations, while the middle equality may be attributed to the closure of $\varphi$ and the homology condition. We call a closed $p$-form $\varphi$ satisfying (\ref{cal1}) a \textit{calibration} and the submanifold $\mathcal{N}$ to be a \textit{calibrated cycle} in a \textit{calibrated manifold} $\mathcal{M}$.  

The simplest example of a calibrated manifold one may consider is a complex manifold with real dimension $2n$ and a K\"{a}hler form $J$ and $\varphi = \frac{1}{p!} J^{p}$ with $ 1\leq p \leq n$. If $\dd \varphi =  0$ so that $\mathcal{M}$ is K\"{a}hler, then the submanifolds (cycles) calibrated by $\varphi$ are homologically volume minimising and may be referred to as K\"{a}hler cycles. As we restrict ourselves to 3D with a constant time condition, the resulting 2D Riemannian manifold must be K\"{a}hler, so the only K\"{a}hler cycle is calibrated by the volume form. 

This motivates us to consider sLag cycles, which are calibrated by the real part of the holomorphic one-form $\varphi = e^{i \chi} \Omega$, where $\chi$ is an arbitrary phase. In general, a submanifold $\mathcal{N}$ is sLag if and only if the following two conditions are satisfied \cite{Harvey:1982xk}: 
\begin{align} 
J |_{\mathcal{N}} &= 0, \\
\textrm{Im} (\varphi) |_{\mathcal{N}} &= 0.  
\end{align}
In other words, we must ensure that the K\"{a}hler form and the imaginary part of the holomorphic form vanish when restricted to the sLag submanifold. Given we are working in 2D, the first condition is trivial, while the second along with the closure of the real part of $\varphi$, $\dd \textrm{Re}(\varphi) = 0$, needs to be imposed to ensure that the submanifold or cycle is volume minimising. 

In order to define a generalised   calibration \cite{Gutowski:1999iu, Gutowski:1999tu}, let us introduce a potential energy functional, or action, of the form 
\be
\label{gen_cal_action}
S = \int \dd^p \sigma [ \sqrt{\textrm{det} \, g} + A ]
\ee
where $g$ is the induced worldspace $p$-metric, $\sigma^i, i = 1, \dots, p$ denote spatial coordinates and $A$ is a $p$-form potential with field strength $F= \dd A$. Let us again consider $\mathcal{N}$ a submanifold of a calibrated manifold with generalised  calibration $\varphi$, so that (\ref{cal2}) holds. Let $\mathcal{N}'$ be a submanifold  that is in the same homology class as $\mathcal{N}$ with $\partial \mathcal{N} = \partial \mathcal{N}'$. Then we can apply the same argument as above: we note that 
\be
\vol(\mathcal{N}) = \int_{\mathcal{N}} \varphi = \int_{\mathcal{N}'} \varphi + \int_{D} \dd \varphi, 
\ee
where $D$ is a $(p+1)$-dimensional surface with $\partial D = \mathcal{N} - \mathcal{N}'$. Now, provided 
\be
\dd \varphi = -F, 
\ee
then we have 
\be
\int_{\mathcal{N}} \dd^{p} \sigma [ \sqrt{\textrm{det} \, g} + A ] = \int_{\mathcal{N}'} \varphi + \int_{\mathcal{N}'} \dd^p \sigma A 
\leq \int_{\mathcal{N}'} \dd^{p} \sigma [ \sqrt{\textrm{det} \, g} + A ],   
\ee
where the final inequality follows from (\ref{cal1}). As a result of this argument, we conclude that $S|_{\mathcal{N}} \leq S|_{\mathcal{N}'}$, so that the action restricted to the submanifold $\mathcal{N}$ is minimised. It is easy to see that if $F = 0$, then the bound for generalised calibrations reduces to the bound for calibrated manifolds saturated by minimal surfaces.

Before leaving this section, one final important comment is in order. It is known that (generalised) calibrations are intimately related to supersymmetry conditions that follow from the Killing spinor equation. For example, in  \cite{Gauntlett:2006ai} supersymmetric M2-branes are considered and one can define a one-form $K$ and two-form $\Sigma$ from the Killing spinor bilinears and show that $\dd \Sigma = i_{K} G_4$, where $G_{4}$ is the four-form flux of 11D supergravity. In this process, one identifies $\Sigma$ as a generalised   calibration. With this connection in mind, we would like to see if one can define one-form Killing spinor bilinears in AdS$_3$ that play the same role. 

From the perspective of supersymmetry with Killing spinor $\epsilon$, it is most natural to define Killing spinor vector bilinears 
\be
\begin{aligned}
K_{\mu} &= i \bar{\epsilon} \gamma_{\mu} \epsilon = i \epsilon^{\dagger} A \gamma_{\mu} \epsilon, \\
\Omega_{\mu} &= \bar{\epsilon}^{c} \gamma_{\mu} \epsilon = - \epsilon^{T} C^{-1} \gamma_{\mu} \epsilon. 
\end{aligned}
\ee
where in 3D with signature $(-,+,+)$ $A$ and $C$ satisfy $A \gamma_{\mu} A^{-1} = - \gamma_{\mu}^{\dagger}$ and $C^{-1} \gamma_{\mu} C = - \gamma_{\mu}^T$. We observe that $K$ is a real one-form, whereas $\Omega$ is complex, so we have precisely enough vectors to define a 3D spacetime. 

Let us be more specific and consider massless BTZ (\ref{massless_BTZ}), where the solution to the Killing spinor equation, $\nabla_{\mu} \epsilon = \frac{1}{2} \gamma_{\mu} \epsilon$, is 
\be
\label{epsilon}
\epsilon =( r^{\frac{1}{2}}  + r^{-\frac{1}{2}} x \gamma_{x} ) \epsilon_+  + r^{-\frac{1}{2}} \epsilon_-, \quad \gamma_{r} \epsilon_{\pm} = \pm \epsilon_{\pm}. 
\ee
Since we are interested in constant time surfaces, we have set $t$ to a constant, which allows us without loss of generality to absorb it in the constant spinors $\epsilon_{\pm}$. In \cite{Colgain:2016lxi}, it was noted that the RT embedding (\ref{RT_massless}) preserved half the Killing spinors provided the constant spinors satisfied the relation,    
\be
\label{constraint}
\epsilon_- = h \gamma_{x} \epsilon_+,  
\ee
where $h$ is a constant. Substituting these expressions back into $K$ and $\Omega$, while neglecting $e^{0} = \frac{\dd t}{r}$, since $t$ is constant, but retaining $e^{x} = \frac{\dd x}{r}, e^{r} = \frac{\dd r}{r}$ components, it is easy to explicitly check that both one-forms vanish on the minimal surface. The vanishing of $K$ is not surprising: it defines a timelike vector and evaluated on a spacelike surface, such as the RT surface, it can be expected to be zero, since it is normal to the surface. In contrast, the vanishing of $\Omega$ is a little puzzling, since this has precisely the right form to define a sLag cycle. 

However, it is easy to understand this from another angle to confirm that it must vanish. Let us define the additional scalar $f = \bar{\epsilon} \epsilon$. Using Fierz identity, one can show that supersymmetry restricts the norms of $K$ and $\Omega$ to satisfy: 
\be
-K^2 = \textrm{Re}(\Omega)^2 = \textrm{Im}(\Omega)^2 = f^2. 
\ee
As claimed earlier, it is easy to see that if $f$ is non-zero, then $K$ defines a timelike direction. However, once we consider constant time surfaces, then $K=0$ implies $f=0$. As a result, we are left with $\Omega$, which is defined on a Riemannian space and has zero norm, which implies it is also zero.  

So, the take-home message is that while one can define spinor bilinears, at least in 3D in the case of locally AdS$_3$ solutions, these bilinears cannot correspond to calibrations for spacelike surfaces. This appears to preclude the possibility that we can use supersymmetry conditions derived from the Killing spinors to identify calibrations that can be applied to calculate holographic entanglement entropy. 

\section{Calibrations and BTZ black holes}
\label{sec:BTZ}
Having introduced calibrations, in this section we illustrate their utility in the context of the class of locally AdS$_3$ spacetimes corresponding to BTZ black holes \cite{Banados:1992wn, Banados:1992gq}. We emphasise that the same results may be achieved from solving the geodesic equation, which is easy to do in 3D since all BTZ black holes have a global $U(1) \times U(1)$ isometry that allows one to introduce two constants of motion. Moreover, the same outcome is achieved by studying supersymmetric curves \cite{Colgain:2016lxi}, but as we remarked in the last section, it is not immediately obvious how calibrations and supersymmetry are reconciled in the current context. 

As stressed in the introduction, calibrations allow us to by-pass the second-order equations and reduce the problem immediately to solving first-order partial differential equations (PDEs). We will in turn solve the latter using the method of characteristics (see for example \cite{Courant_Hilbert}). In order to help the reader digest the method, we start by considering the simplest case of massless BTZ, before proceeding to static BTZ and non-extremal, rotating BTZ. 

\subsection{Massless BTZ}
We begin with the simplest example, namely massless BTZ in Poincar\'e patch, where the spacetime metric is
\be
\label{massless_BTZ}
\dd s_3^2 = \frac{1}{r^2} ( - \dd t^2 + \dd x^2 + \dd r^2). 
\ee
Here, it is easy to identify a spacelike hypersurface by adopting constant time. The 2D space is then hyperbolic and one can introduce the holomorphic form 
\be
\varphi = e^{i \chi} \left( \frac{\dd x}{r} + i \frac{\dd r}{r} \right). 
\ee
Modulo an ambiguity in the phase $\chi$, this is then our candidate sLag calibration. To guarantee it is genuinely sLag, we must ensure that the imaginary part vanishes and the real part is closed. This leads us to the two equations:  
\begin{align}
\label{massless_cal1} \cos \chi \frac{\dd r}{r} + \sin \chi \frac{\dd x}{r}  &= 0, \\
\label{massless_cal2} \partial_r \left( \frac{\cos \chi}{r} \right) + \frac{1}{r} \partial_{x} (  \sin \chi) &= 0. 
\end{align}
The philosophy now is to solve (\ref{massless_cal2}) for $\chi(r, x)$ before substituting back into (\ref{massless_cal1}). 

The solution to (\ref{massless_cal2}) can be found using the method of characteristics as we summarise now. First, let us denote $\cos\chi = f(x,r)$ for convenience. The equation~\eqref{massless_cal2} is then
\be\label{pde2}
		\frac1r \,\frac{\partial f}{\partial r} - \frac{f}{r \sqrt{1-f^2}} \,\frac{\partial f}{\partial x} = \frac{f}{r^2}.
\ee
In an auxiliary three-dimensional space spanned by $(r,x,f)$ this equation can be viewed as a condition of orthogonality between the two vector fields. One of them can be read off from~\eqref{pde2}:
	\be
		V^i = (V^r, V^x, V^f) = \left(\frac1r, - \frac{f}{r \sqrt{1-f^2}},\frac{f}{r^2}\right),
	\ee
while the other is a field of normal vector $N^i$ to a two-dimensional surface given by the equation $f=f(r,x)$:
	\be
		N^i = (N^r, N^x, N^f) = \left(\frac{\partial f}{\partial r}, \frac{\partial f}{\partial x}, -1\right).
	\ee
Thus the equation~\eqref{pde2} can be viewed as a condition that the surface $f=f(r,x)$ is a one-parameter family of integral curves of $V^i$ (the normal vector to the surface is orthogonal to $V^i$). To find the integral curves, write down the system of ordinary differential equations (ODEs):
	\be
		\dot r = V^r,\qquad \dot x = V^x,\qquad \dot f = V^f,
	\ee
or equivalently 	
	\be\label{ode1}
		r \, \dd r = -\frac{r \sqrt{1-f^2}}{f} \,\dd x = \frac{r^2}{f}\,\dd f.
	\ee
It is worth noting that the first equality is simply the equation (\ref{massless_cal1}), so by solving via the method of characteristics, the remaining calibration condition is guaranteed to hold. 

As soon as we find two independent first integrals of this system
	\be
	\begin{aligned}
		c_1 &= \phi_1 (f,x,r),\\
		c_2 &= \phi_2 (f,x,r),
	\end{aligned}
	\ee
any integral curve of $V^i$ corresponds to some fixed values of $c_1$ and $c_2$. A one-parameter family of integral curves, then, is given by fixing some functional dependence $F(c_1,c_2)=0$. Any choice of function $F$ gives some integral surface of the vector field $V^i$, and if one is able to solve $F(c_1,c_2)=0$ for $f$, this would give some solution $f=f(r,x)$ to~\eqref{pde2}.	

In particular, from~\eqref{ode1} we see that
	\be
		r\,\dd r = \frac{r^2}{f}\,\dd f,
	\ee
which immediately implies that $f = c_1 r$. Substituting for $r$, we can recast the remaining equation from~\eqref{ode1} as
	\be
		-c_1 \,\dd x = \frac{f\,\dd f}{\sqrt{1-f^2}},
	\ee	
which can be integrated to give $-c_1 x + \sqrt{1-f^2} = c_2$. We remark that at this stage we could employ the shift symmetry available in $x$ to set $c_2 =0$. However, for the moment we retain it. Thus, the two first integrals of the system~\eqref{ode1} are given by
	\be\label{int}
	\begin{aligned}
		c_1 &= \frac fr,\\
		c_2 &= \sqrt{1-f^2} - f\,\frac xr.
	\end{aligned}
	\ee
Note, these are independent first integrals of the system of ODEs and at the same time implicit solutions to the PDE. More generally, one could proceed by choosing various functions $F(c_1,c_2)=0$ to derive a generic solution $f=f(x,r)$ to~\eqref{pde2}. However, let us look at the characteristics themselves. A characteristic is an integral curve of $V^i$, given by intersection of the surfaces~\eqref{int}. It is easy to exclude $f$ from these algebraic equations, which gives a projection of an arbitrary characteristic to the $(x,r)$ plane:
	\be
	\label{eliminating_f}
	\left( x + \frac{c_2}{c_1} \right)^2 + r^2 = \frac{1}{c_1^2}. 
	\ee

It is worth stressing that $c_1$ is a constant of motion that arises from the fact that $x$ is an isometry direction. Relabeling it as $c_1 = h^{-1}$, while employing shift symmetry in $x$ to set $c_2 =0$, we arrive at the known RT minimal surface for massless BTZ, 
\be
\label{RT_massless}
r^2 + x^2 = h^2.  
\ee
 
Note, in solving the calibration conditions, we have not extracted an expression for $f$, or alternatively $\chi$. Indeed, we have identified two integral surfaces of the characteristic vector field (\ref{int}) and we should make sure that the $\textrm{Re}( \varphi)$ agrees with $\dd s$, the differential of length of the geodesic, on their intersection, so that we recover the usual RT prescription. This ensures the validity of equation (\ref{SEE}). For the first surface in (\ref{int}), with $c_1 = h^{-1}$, the corresponding calibration is 
\be
\textrm{Re} (\varphi) = \frac{1}{h} \left( \dd x- \frac{x}{r} \dd r \right)  = \dd s, 
\ee
where we have used (\ref{RT_massless}) to simplify the expression. Repeating the exercise with the second integral surface, once again employing (\ref{RT_massless}), this time with $c_2 = 0$, we find the same result.  This is guaranteed to be the case since (\ref{eliminating_f}), from where we deduce (\ref{RT_massless}), is simply the intersection of the two surfaces given in (\ref{int}). 

\subsection{Static BTZ}

Having mastered the simplest case, we move onto the static BTZ black hole with spacetime metric, 
\be
\label{static_BTZ}
\dd s_3^2 = - (r^2-m) \dd t^2 + r^2 \dd x^2 + \frac{\dd r^2}{(r^2-m)}. 
\ee
We again restrict our attention to a constant time hypersurface, and to simplify expressions, we redefine $r = \sqrt{m} \cosh \rho$, so that the 2D hypersurface metric becomes: 
\be\label{rho-metric}
\dd s_2^2  = m \cosh^2 \rho \, \dd x^2 + \dd \rho^2. 
\ee 
We next introduce the holomorphic one-form, 
\be
\varphi= e^{i \chi} (\sqrt{m} \cosh \rho\, \dd x + i \dd \rho), 
\ee
which serves as the sLag calibration. We proceed to identify the calibration conditions. Denoting $\cos\chi = f(x,r)$ we recast the $\dd \textrm{Re} ( \varphi) = 0$ equation in the form 
	\be\label{pde3}
	\sqrt{m} \cosh \rho \,\frac{\partial f}{\partial\rho} - \frac{f}{\sqrt{1-f^2}} \,\frac{\partial f}{\partial x} = -\sqrt{m} \,f \sinh\rho.
	\ee
We thus need to look for two independent first integrals of the system of ODEs, which defines the characteristics of~\eqref{pde3}:
	\be\label{ode2}
	\frac{\dd \rho}{\sqrt{m}\cosh\rho} = -\frac{\sqrt{1-f^2}}{f} \,\dd x = -\frac{1}{\sqrt{m}\sinh\rho}\,\frac{\dd f}{f}.
	\ee
We begin by solving the equation that relates $\dd \rho$ and $\dd f$, which yields $c_1 = f \cosh \rho$. Using this to eliminate $f$ from the equation that relates $\dd \rho$ and $\dd x$, we get the $\textrm{Im}(\varphi) = 0$ condition. This again shows that the method of characteristics takes care of the other calibration condition. On the other hand, eliminating $\rho$ from the equation that relates $\dd x$ and $\dd f$,  we find
\be
\label{integral0}
\sqrt{m} \, \dd x = \frac{f \, \dd f}{\sqrt{(1-f^2)(c_1^2-f^2)}}. 
\ee
As we show in the appendix for the rotating case, if one demands that the minimal surface makes contact with the boundary to define a spacelike separated interval, then we require $c_1^2 > 1$. We postpone motivating this condition further until we discuss the rotating BTZ example in the next subsection.
 
In integrating (\ref{integral0}), one has to exercise some care in order to find the correct characteristics. Denoting $y=f^2$, $a= \frac12(c_1^2+1)$, $b= \frac12(c_1^2-1)$, we have
	\be\label{integral}
	2\sqrt{m}(x-c_2) = \int \frac{dy}{\sqrt{(y-a)^2-b^2}} = -t,
	\ee
where we have used the substitution 
\be
\cosh t = \frac{a-y}{b}, 
\ee
where we are assuming $a -y > b > 0$. Using $c_1 = f\cosh\rho$ and simplifying, one can solve for $f$:
	\be
	\label{coth}
	f = \left[1+\sinh^2\rho\, \tanh^2 (\sqrt{m}x-c_2)\right]^{-1/2}. 
	\ee
Together with
	\be\label{cosh}
	f = \frac{c_1}{\cosh\rho}, 
	\ee
we have two families of solutions to~\eqref{pde3} that represent different integral surfaces of the characteristic vector field. The families are parametrised by the values of the first integrals, $c_1, c_2$, of the system of ODEs~\eqref{ode2}.

\begin{figure}[t]
	\begin{center}
		\includegraphics[scale=0.37]{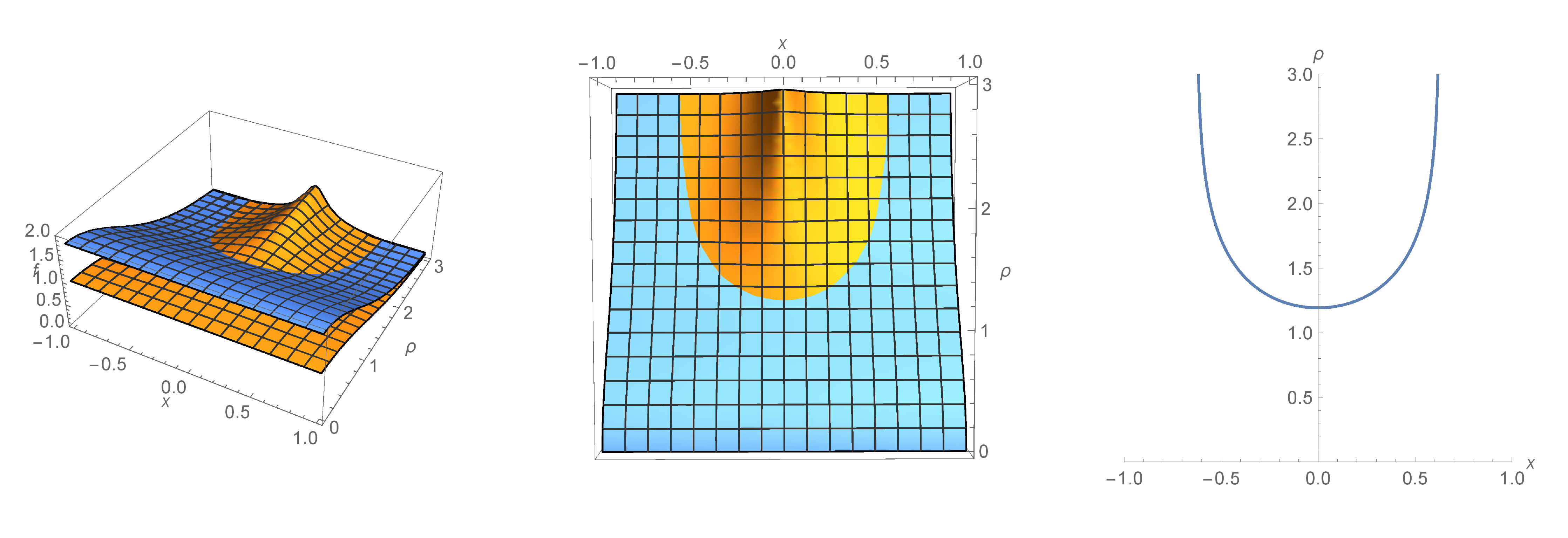}
		\caption{Surfaces~\eqref{coth} with $c_2 = 0, m = 1$, orange and~\eqref{cosh} with $c_1 = 1.8$, blue (left); the same surfaces viewed from above (centre); the curve~\eqref{geod-stat} (right).}
		\label{picture}
	\end{center}
\end{figure}

The intersection between the integral surfaces~\eqref{coth} and~\eqref{cosh} for given values of $c_1, c_2$  is the characteristic of the initial PDE~\eqref{pde3}. By eliminating $f$ we find the projection of the characteristic onto the $(x,\rho)$ plane:
	\be\label{geod-stat}
	\tanh\rho =\sqrt{\frac{c_1^2-1}{c_1^2}} \cosh \sqrt{m}(x-c_2)
	\ee
Now we can fix $c_1 = r_*/\sqrt{m}$ and set $c_2 = 0$ by employing a shift in $x$. In terms of the original coordinates, the curve then becomes 	
\be
	\label{RT_static_BTZ}
	\frac{\sqrt{r^2-m}}{r} = \frac{\sqrt{r_*^2-m}}{r_*} \cosh \sqrt{m} x, 
	\ee
where it can be confirmed from equation (4.25) of \cite{Colgain:2016lxi} that this is the expected RT minimal surface. At this point, we should check that $\textrm{Re} (\varphi)$ agrees with $\dd s$ to make sure that everything is consistent. It can be verified that the two first integrals of the system of ODEs agree, as expected, so we simply choose $f = \cos \chi = r_*/(\sqrt{m} \cosh \rho)$. A straightforward calculation then reveals that 
\be
\textrm{Re} ( \varphi) = r_* \dd x - \frac{\sqrt{r_*^2 -m}}{\sqrt{r^2-m}} \sinh (\sqrt{m} x) \dd r = \dd s. 
\ee

\subsection{Rotating BTZ}
This brings us to our last example, namely (non-extremal) rotating BTZ:
\be
\label{rot_BTZ}
\dd s_3^2  = - \frac{(r^2 - r_+^2)(r^2 - r_-^2)}{r^2} \dd t^2 + \frac{r^2 \dd r^2}{(r^2 - r_+^2)(r^2 - r_-^2)} + r^2 \left( \dd x + \frac{r_+ r_-}{r^2} \dd t \right)^2, 
\ee
where we will be brief and omit details, since they closely mirror the previous example. In order to identify the required 2D spacelike hypersurface, we rewrite (\ref{rot_BTZ}) as 
\be
\label{rotating_BTZ}
\dd s_3^2 = \frac{ - (r^2 - r_+^2) \dd X_-^2 + (r^2 - r_-^2) \dd X_+^2 }{r_+^2-r_-^2} +  \frac{r^2 \dd r^2}{(r^2-r_+^2)(r^2-r_-^2)}, 
\ee
where we have defined $X_{\pm} = r_{\pm} x + r_{\mp} t$. 
We observe that $X_{-}$ is a timelike direction, so we define our spacelike surface by setting it to be a constant. By way of a side remark, it can be checked from the analysis in the appendix that the geodesic equations can be consistently truncated in this fashion, in line with expectations. For simplicity, we choose $X_- = 0$ and drop the remaining subscript. 
To exploit the similarity with the last example, we switch to a new radial coordinate,
\be
	\sqrt{\frac{r^2-r_-^2}{r_+^2-r_-^2}} = \cosh\rho. 
\ee
As a result, the metric on the hypersurface can be simplified accordingly, 
\be
\label{BTZ_hyper}
\dd s_2^2 = \frac{r^2-r_-^2}{r_+^2-r_-^2} \dd X^2 +  \frac{r^2 \dd r^2}{(r^2-r_+^2)(r^2-r_-^2)} = \cosh^2\rho\, \dd X^2 + \dd \rho^2. 
\ee
We observe that this is just the metric~\eqref{rho-metric} up to replacement $m\rightarrow1$, $x\rightarrow X$. Therefore, the subsequent analysis carries over from the static case and we can immediately reproduce the result~\eqref{geod-stat} with the same replacement:
	\be\label{geod-rot}
		\sqrt{\frac{r^2-r_+^2}{r^2-r_-^2}} = \sqrt{\frac{c_1^2-1}{c_1^2}} \cosh (X-c_2). 
	\ee
For the choice $c_1 = L/r_+$ and $c_2=0$, where we have again exploited shift symmetry, we have
\be
\label{RT_rot}
\cosh X =  \frac{L}{\sqrt{L^2 - r_+^2}} \, \sqrt{\frac{r^2-r_+^2}{r^2-r_-^2}}, 
\ee
which recovers equation (4.38) of \cite{Colgain:2016lxi} upon setting $X_- = 0$. Note that $L  \equiv g_{x \mu} \dot{x}^{\mu}$ is a constant of motion associated to the isometry in the original $x$-direction given in (\ref{rot_BTZ}). As a further consistency check, it can be noted that setting $r_- = 0, r_+ = \sqrt{m}$ and $L = r_*$, we recover the earlier expression for static BTZ (\ref{RT_static_BTZ}). Finally, it can be checked that $\textrm{Re} (\varphi) = \dd s$, so that (\ref{SEE}) agrees with the RT prescription. 

We now comment on the restriction $c_1^2 > 1$. In the appendix we have solved the geodesic equation for rotating BTZ metric to make sure that there is well-defined interval on the boundary with spacelike separation. Up to an irrelevant sign, it can be seen from (\ref{L_Delta}) that $ |c_1| = |L/r_+| > 1$, so that only when $c_1^2 > 1$ do we find a good geodesic for the specific purpose of calculating entanglement entropy holographically. 

\section{Generalised  Calibrations and Warped AdS$_3$}
\label{sec:warped}
Having discussed BTZ black holes, which are a class of locally AdS$_3$ solutions, in the last section, here we consider one of the simplest deformations of the AdS$_3$ geometry. We will focus on spacelike warped AdS$_3$ solutions, where the warping is sourced by a $U(1)$ gauge field. While warped AdS$_3$ vacua arise in a host of different settings, including Topologically Massive Gravity \cite{Nutku:1993eb, Gurses:1994bjn, Bouchareb:2007yx}, the near-horizon of extremal Kerr black holes \cite{Bardeen:1999px, Bengtsson:2005zj}, as well as supersymmetric solutions to $\mathcal{N} = 2$ off-shell supergravities \cite{Deger:2013yla, Deger:2016vrn}, here we confine our attention to the following theory \cite{Detournay:2012dz},   
\be
\begin{aligned}
\mathcal{L} &=  R \, \vol_3 - 4 \dd U \wedge *_3 \dd U - 4 e^{-4 U} A \wedge *_3 A \nn 
&+ 2 e^{-4 U} (2 - e^{-4 U} ) \vol_3 - A \wedge F, 
\end{aligned}
\ee
where $A$ is a $U(1)$ gauge field with field strength $F= \dd A$ and $U$ denotes a scalar. This theory can be defined as a consistent truncation of 10D supergravity \cite{Detournay:2012dz}. Moreover, with the inclusion of some additional fields, it can be brought to the form of a 3D $\mathcal{N}=2$ gauged supergravity \cite{Karndumri:2013dca}, once again embedded in 10D supergravity. The advantage of focusing on this theory is that the dual theory is believed to be a 2D CFT, since one can recover two copies of the Virasoro algebra from the asymptotic symmetry analysis \cite{Compere:2014bia}. For this reason, we can view it as one of the mildest deformations of AdS$_3$. 

For constant $U$, the theory admits a family of warped black string solutions, which are parametrised by left/right-moving temperatures $T_{\mp}$ and an arbitrary parameter $\lambda$ \cite{Compere:2014bia} 
\be
\begin{aligned}
\label{warped_metric}
\dd s_3^2 &=  T_+^2 \dd v^2 + 2 \rho\, \dd u \, \dd v + \left[ T_-^2 e^{4 U}- \lambda^2 \rho^2 \right] \dd u^2 
+ \frac{ e^{4 U} \dd \rho^2}{4 (\rho^2 - T_+^2 T_-^2)}, \\
e^{4 U} &= 1 + \lambda^2 T_+^2, \quad A =  \lambda \, e^{-2 U} ( T_+^2 \dd v + \rho \, \dd u). 
\end{aligned}
\ee
One of the key observations of \cite{Compere:2014bia} is that for fixed $U$ one can define an auxiliary \textit{unwarped} AdS$_3$ metric $\tilde{g}_{\mu \nu}$, which is related to the warped metric $g_{\mu \nu}$: 
\be
\tilde{g}_{\mu \nu} = e^{-4 U} g_{\mu \nu} + A_{\mu} A_{\nu}.  
\ee
Restricting our attention to the above solution, the explicit unwarped metric may be expressed as \cite{Compere:2014bia}, 
\be
\label{auxiliary_metric}
\dd s_3^2 = T_+^2 \dd v^2 + 2 \rho \, \dd u \,  \dd v + T^2_- \dd u^2 + \frac{\dd \rho^2}{4 (\rho^2 - T_+^2 T_-^2)}. 
\ee
Although this may look unfamiliar, it is an easy exercise to recast the above metric as rotating BTZ (\ref{rot_BTZ}) through the following redefinitions:  
\be
\label{transform}
v = \frac{1}{\sqrt{2}} (- t + x), \quad u = \frac{1}{\sqrt{2}} ( t + x),  \quad \rho = r^2 - \frac{1}{2} ( r_+^2 + r_-^2), \quad T_{\pm} = \frac{1}{\sqrt{2}} ( r_{+} \mp r_{-} ). 
\ee

At this point, we import a key result from Song et. al \cite{Song:2016pwx}. As remarked earlier, it is straightforward to determine holographic entanglement entropy in AdS$_3$ as the problem reduces to calculating the length of spacelike geodesics. While this is true in the auxiliary unwarped AdS$_3$ geometry, it is not true in the warped counterpart. In fact, in the warped geometry one must consider the trajectory of a charged particle. To illustrate the distinction, let us consider the action 
\be
\label{warped_action}
S = \frac{1}{4 G_3} \int  \dd s \left[  m \sqrt{ g_{\mu \nu} \dot{x}^{\mu} \dot{x}^{\nu}} + q  A_{\mu} \dot{x}^{\mu} \right], 
\ee
where $m$ is the mass and $q$ is the charge. Now, provided the constants are chosen, such that 
\be
\label{qm}
\frac{q}{m} = A_{\mu} \dot{x}^{\mu} e^{4 U}, 
\ee
and one normalises the velocity of the particle so that $g_{\mu \nu} \dot{x}^{\mu} \dot{x}^{\nu} =1$, then one recovers the same equations as the geodesic equation in auxiliary AdS$_3$ \cite{Song:2016pwx}. More concretely, one finds that the equation,
\be
\ddot{x}^{\mu} + \Gamma^{\mu}_{\rho \sigma} \dot{x}^{\rho} \dot{x}^{\sigma} = \frac{q}{m} F^{\mu}_{~\nu} \dot{x}^{\nu}
\ee
which follows from the action (\ref{warped_action}), when evaluated on the warped solution (\ref{warped_metric}), agrees with the geodesic equation for auxiliary AdS$_3$ (\ref{auxiliary_metric}), 
\be
\ddot{\tilde{x}}^{\mu} +  \tilde{\Gamma}^{\mu}_{\rho \sigma} \dot{\tilde{x}}^{\rho} \dot{\tilde{x}}^{\sigma}  = 0. 
\ee
Therefore, the problem of finding the trajectory boils down to solving the geodesic equation in auxiliary AdS$_3$. Note, some care is required with the normalisation of the velocity as $g_{\mu \nu} \dot{x}^{\mu} \dot{x}^{\nu} = 1$ implies $\tilde{g}_{\mu \nu} \dot{\tilde{x}}^{\mu} \dot{\tilde{x}}^{\nu} \neq 1$. 

One further comment: from (\ref{qm}) it is not immediately obvious that the right hand side is a constant. To see this, note that the only raised component of $A_{\mu}$, namely $A^{v} = \lambda ( 1 + \lambda^2 T_+^2)^{-\frac{1}{2}} \partial_{v}$, is a Killing vector, so that there is a constant of motion associated to it. This property ensures that the right hand side is a constant. 

\subsection{Warped geometry}
In principle we could use of the method of characteristics introduced earlier to solve the generalised  calibration conditions for the warped geometry. However, we have already extracted an expression for the sLag cycle by solving the calibration condition in the unwarped auxiliary AdS$_3$, which through the coordinate transformation (\ref{transform}) may be brought to the form of rotating BTZ (\ref{rot_BTZ}). This reduces the problem to the analysis presented in section 3. For this reason, here we will simply confirm that the warped geometry satisfies a generalised  calibration condition. 
 
As before, let us consider $X_- = 0, X_+ = X$. From the perspective of the warped geometry (\ref{warped_metric}) it is not immediately obvious how to select the 2D hypersurface, but here we can use the existence of the auxiliary AdS$_3$ to guide us. With this simplification, we can rewrite the warped AdS$_3$ coordinates $(v, u, \rho)$ in terms of $(X, r)$, 
\be
v = \frac{1}{\sqrt{2}} \frac{X}{(r_+ - r_-)}, \quad u = \frac{1}{\sqrt{2}} \frac{X}{(r_+ + r_-)}, \quad \rho = r^2 - \frac{1}{2} ( r_+^2 + r_-^2), 
\ee
where now $r$ is the radial coordinate of the BTZ metric. Following the same steps as before, we isolate the 2D spacelike hypersurface,  
\be
\dd s^2_2 = \frac{(r^2 - r_-^2)}{(r_+^2- r_-^2)} \Delta_1 \dd X^2 + \frac{ r^2 \, \Delta_2}{(r^2- r_+^2)(r^2 - r_-^2)} \dd r^2, 
\ee
where we have defined 
\be
\Delta_1 =  1 - \frac{\lambda^2}{2} \frac{(r_+ - r_-)}{(r_+ + r_-)} (r^2 - r_+^2), \quad \Delta_2 = 1 + \frac{\lambda^2}{2} (r_+ - r_-)^2. 
\ee
Setting $\lambda = 0$, it is easy to check that we recover the unwarped 2D hypersurface (\ref{BTZ_hyper}). 

With this 2D metric, the candidate calibration becomes, 
\be
\label{gen_cal0}
\varphi = e^{i \chi} \left(  \sqrt{\frac{r^2 - r_-^2}{r_+^2 - r_-^2}} \sqrt{\Delta_1} \dd X + i \sqrt{\Delta_2} \frac{r \dd r}{\sqrt{(r^2 - r_+^2)(r^2-r_-^2)}} \right). 
\ee
To show that this is a generalised  calibration, we require that its imaginary part vanishes and that 
\be
\label{gen_cal}
\dd \textrm{Re} (\varphi) = - \frac{q}{m} F, 
\ee
where the constant factor is fixed by comparing the warped action (\ref{warped_action}) with (\ref{gen_cal_action}). 

We begin by determining the right hand side in terms of $\dot{X}$,
\be
\frac{q}{m} F_{r X}   = \frac{\lambda^2 \dot{X} ( r^2 - r_-^2) r}{( r_+ + r_-)^2 }. 
\ee
where we have reverted to coordinates. At this point, we should ensure that $\dot{X}$ is correctly normalised so that $g_{\mu \nu} \dot{x}^{\mu} \dot{x}^{\nu} = 1$ in the warped metric. To determine this, one can use (\ref{RT_rot}) to eliminate $\dot{r}$ in terms of $\dot{X}$, so that one can solve $g_{\mu \nu} \dot{x}^{\mu} \dot{x}^{\nu} = 1$ for $\dot{X}$. This gives a final expression for the field strength: 
\be
\frac{q}{m} F_{r X} =  -  \frac{\sqrt{2} \lambda^2 L  (r_+ - r_-) r }{(r_+ + r_-) \sqrt{ (L^2 - r_+^2)[1 + (r_+ -r_-)^2 \lambda^2 ] -(L^2 + r_+^2)}}.
\ee

It is easy to see that one recovers, up to sign, the same expression from $\dd\textrm{Re} ( \varphi)$. To do, so we first use the vanishing of the imaginary part of the calibration, along with (\ref{RT_rot}) to determine $\tan \chi$, 
\be
\tan \chi  = - \frac{\sqrt{\Delta_2} \sqrt{(r^2 - r_+^2) (L^2 + r_+^2)- (r^2 - 2 r_-^2 + r_+^2) (L^2 - r_+^2)}}{\sqrt{2} L  \sqrt{\Delta_1} \sqrt{r_+^2 - r_-^2}}, 
\ee
which in turn allows us to extract expressions for $\cos \chi $ and $\sin \chi$. It is worth noting that $\chi$ only depends on the radial direction $r$. This means that substituting back into the calibration (\ref{gen_cal0}), we only need to consider the derivative of the first term. Performing this step, and simplifying accordingly, one indeed confirms that (\ref{gen_cal}) is satisfied. This confirms that the sLag cycle we have identified is calibrated with respect to a generalised calibration. 

\section{Calibrations in warped AdS$_3$}
\label{sec:warped_geo}
We claim that calibrations offer a unified way to determine holographic entanglement entropy. For completeness, in this section we will demonstrate the utility of calibrations in identifying \textit{geodesics} in warped AdS$_3$ black holes. Similar analysis for warped AdS$_3$ spacetimes without a horizon have appeared in \cite{Anninos:2013nja}. It is worth noting that spacelike geodesics in warped AdS$_3$ have different asymptotics to unwarped AdS$_3$: instead of a constant interval at the boundary, we will see that the interval is infinite, a feature noted earlier in \cite{Anninos:2013nja}.   

We recall the metric ({\ref{warped_metric}}), in the generic case for which $T_+ \neq 0$ and $T_- \neq 0$, and make the coordinate transformation 
\be
u = {1 \over T_-} {\hat{u}}, \quad v={1 \over T_+} {\hat{v}},
\quad \rho= T_+ T_- {r}, 
\ee
and further define
\be
\mu^2=1+\lambda^2 (T_+)^2 \ .
\ee
On performing the coordinate transformation, and dropping the hats,
the metric simplifies to
\be
\label{msimp1}
\dd s_3^2 = -\mu^2 (r^2-1)\dd u^2 +(\dd v+r \, \dd u)^2+{\mu^2 \over 4(r^2-1)} \dd r^2 \ .
\ee

To initiate our analysis, we make a change of coordinates to identify a preferred timelike direction, 
\be
u = \frac{1}{2} (x-t), \quad v = \frac{1}{2} (t +x). 
\ee
In these new coordinates, 
the metric is expressed as 
\be
\label{3d_metric}
\dd s^2_3 = \frac{1}{4} \left( (1-r)^2 + \mu^2 (1-r^2) \right) \left(\dd t - \frac{(r+1)(\mu^2-1)}{[(1-r)+\mu^2 (1+r)]} \dd x \right)^2 + \dd s^2_2, 
\ee
where we have defined, 
\be
\dd s^2_2 = \frac{\mu^2 (1+r)}{1+ \mu^2 + r (\mu^2-1)} \dd x^2 + \frac{\mu^2}{4 (r^2-1)} \dd r^2.  
\ee
We note that there is a horizon at $r=1$, so our task is to identify a minimal surface that makes contact with the boundary (large $r$), but does not penetrate the horizon. 

Following the procedure outlined for BTZ black holes in section \ref{sec:BTZ}, our next step is to set the timelike direction to zero. Here, the identification of an appropriate timelike direction has been guided by a study of the constants of motion. It is an interesting feature of our choice that both the constants of motion become equal and the change of the $t$-direction along the resulting curve is independent of the radial direction, 
\be
\dot{t} = \frac{\alpha}{\mu^2} ( \mu^2-1), 
\ee
where $\alpha$ is the constant of motion. Again, we draw the attention of the reader to the simplification that results in the unwarped case, $\mu=1$, where the curve is independent of $t$. 

We can now introduce the calibration. In terms of the frame, 
\be
{\bf{e}}^1 = \frac{\mu \sqrt{r+1}}{ \sqrt{1 + \mu^2 + r(\mu^2-1)}} \dd x,
\qquad {\bf{e}}^2={\mu \over 2 \sqrt{r^2-1}} \dd r \, ,
\ee
it can be written as, 
\be
\label{geo_calibration}
\varphi = e^{i \chi} ({\bf{e}}^1+i{\bf{e}}^2). 
\ee
From $\textrm{Im}( \varphi) = 0$, we get the condition, 
\be
\label{im_geo}
\frac{\dd r}{\dd x}  = - \frac{2 \tan \chi \sqrt{r+1} \sqrt{r^2-1} }{\sqrt{1+\mu^2 + r (\mu^2-1)}}. 
\ee
In order to determine the second calibration condition, we note the expression for $\textrm{Re}(\varphi)$,
\be
\textrm{Re} (\varphi) =  \frac{\mu \sqrt{r+1} \cos \chi}{ \sqrt{1 + \mu^2 + r(\mu^2-1)}} \dd x -  {\mu \sin \chi \over 2 \sqrt{r^2-1}} \dd r. 
\ee
In principle, one can now use the method of characteristics to solve the PDE that results from the closure of this one-form. Instead, we will employ a short-cut. Based on earlier analysis, it is clear that an angle $\chi$ can be found that only depends on the radial direction. This assumption, along with $\dd \textrm{Re} (\varphi) = 0$, leads immediately to 
\be
\cos \chi = \frac{\alpha}{\mu} \frac{\sqrt{1+\mu^2 + r(\mu^2-1)}}{\sqrt{r+1}}, 
\ee
where we have fixed the overall constant. Substituting this result back into (\ref{im_geo}), we encounter the differential equation: 
\be
\label{dxdr}
\frac{\dd x}{\dd r} = \pm \frac{\alpha (1+ \mu^2 + r(\mu^2-1))}{2 \sqrt{\alpha^2 (r-1) - \mu^2 (\alpha^2-1) (r+1)} \sqrt{r+1} \sqrt{r^2-1}}, 
\ee
where, assuming $\alpha > 0$, we have allowed for $x$ to increase/decrease with $r$, and care must be taken to ensure that various quantities in square roots are positive: for example, we require, 
\be
\Delta \equiv \mu^2 + \alpha^2 - \alpha^2 \mu^2 >  0. 
\ee
Modulo the sign, a solution to the differential equation (\ref{dxdr}) can be found: 
\bea
x &=& \pm \biggl( \frac{\alpha}{\sqrt{\Delta}} (\mu^2-1) \tanh^{-1} \left[ \frac{\sqrt{(r-1) \alpha^2  - \mu^2 (\alpha^2 -1) (r+1)}}{\sqrt{\Delta} \sqrt{r-1}}\right]  \nn
&+& \tanh^{-1} \left[ \frac{\sqrt{(r-1) \alpha^2  - \mu^2 (\alpha^2 -1) (r+1)}}{\alpha \sqrt{r-1}} \right] \biggr).
\eea
In summary, we have identified the appropriate minimal surface by imposing a calibration condition and it is a straightforward exercise to see that the result corresponds to a solution to the geodesic equation. We have checked that a range of parameters can be found with $\alpha > 1$ where the minimal surface makes contact with the boundary at two points but does not cross the horizon.

\section{Discussion}
In this note we have taken initial steps in applying calibrations to the problem of determining holographic entanglement entropy. This approach may be hoped to reap some benefit in higher dimensions, where the task of identifying minimal surfaces intrinsic to the RT prescription involves solving tricky second-order equations. Here, the rational for using calibrations is that the second-order equations are immediately reduced to first-order PDEs. Furthermore, calibrations provide a more elegant coordinate-free description. Within the scope of this work, we have confined ourselves to calculations in 3D gravity, where the calibration conditions are expected to be equivalent to the geodesic equation. 

For BTZ black holes, we showed that the minimal curves correspond to sLag cycles, which are calibrated by the real part of a holomorphic one-form on a 2D spacelike hypersurface. This allowed us to immediately write down first-order PDEs, which we in turn solved using the method of characteristics. We noted that in the presence of flux, where the spacetime becomes warped AdS$_3$ the sLag cycle is specified by a generalised  calibration, whose exterior derivative is proportional to the flux warping the geometry. Moreover, we showed that calibrations may be employed to identify geodesics in warped AdS$_3$. Thus, calibrations provide a unified approach to determine holographic entanglement entropy in both locally AdS$_3$ and warped AdS$_3$ spacetimes. For warped AdS$_3$ spacetimes that preserve some supersymmetry, it should be possible to identify a projection condition that specifies the required minimal surfaces, thereby generalising the analysis of \cite{Colgain:2016lxi} beyond locally AdS$_3$ spacetimes.

We end with some discussion of the applications to higher-dimensional AdS$_{p+2}$ spacetimes. From the outset, one necessary comment is that the sLag cycle does not generalise in a naive way. To see this, note that in 5D, where $p=3$, the spacelike hypersurface is 4D, implying that the natural sLag cycle is a 2D submanifold. Instead, the RT prescription requires a co-dimension two surface in 5D, or a 3D submanifold, so it is clear that sLag cycles are just unique to 3D.

Regardless, let us consider AdS$_{p+2}$ spacetime with symmetric entangling surfaces corresponding to infinite strips or disks on the boundary. In both cases, it is possible to follow one's nose and identify $p$-forms analogous to the sLag calibration we identified in 3D. These forms may be expressed as  
\be
\label{strip}
\varphi_{\textrm{strip}} = e^{i \chi} \frac{1}{r^{p-1}} \left( \frac{\dd x_1}{r} + i \frac{\dd r}{r} \right) \wedge \dd x_2 \wedge \dots \wedge \dd x_{p}, 
\ee
and
\be
\label{disk}
\varphi_{\textrm{disk}} = e^{i \chi}  \left( \frac{\eta}{r} \right)^{p -1}  \left( \frac{\dd \eta}{r} + i \frac{\dd r}{r} \right) \wedge \vol( S^{p -1}), 
\ee
respectively, where once again $\chi$ is a phase to be determined. In the first case, we have parametrised the metric on $\mathbb{R}^p$ as $\dd s^2 (\mathbb{R}^p) = \dd x_1^2 + \dots + \dd x_p^2$, allowing only $x_1$ to be a function of the radial direction $r$, whereas in the disk case, $\dd s^2 (\mathbb{R}^p) = \dd \eta^2 + \eta^2 \dd s^2 (S^{p-1})$, where only $\eta$ depends on $r$. Owing to the high degree of symmetry, the problem is reduced to the simplicity of 3D. In fact, the above calibrations are sLag cycles in 3D, but not in higher dimensions, since the K\"{a}hler form contracted into $\varphi$ is not zero. However, it is plausible that one can consider the forms to be genuine sLag calibrations in some higher dimensional space by adding extra spectator coordinates that play no role in the analysis. 

In order to demonstrate that the above forms are indeed calibrations, it is enough to show that the conditions $\dd \textrm{Re} (\varphi) = \textrm{Im} (\varphi) = 0$ recover the minimal surfaces identified by Ryu-Takayanagi in higher dimensions \cite{Ryu:2006ef}. For concreteness let us illustrate the case of the disk (\ref{disk}).  Imposing the calibration conditions, we find two equations: 
\bea
0 &=& \sin \chi \dd \eta + \cos \chi \dd r, \nn
0 &=& \partial_{r} \left( \cos \chi \frac{\eta^{p-1}}{r^{p}} \right) + \partial_{\eta} \left( \sin \chi  \frac{\eta^{p-1}}{r^{p}} \right). 
\eea
It is easy to confirm that the higher-dimensional RT minimal surface \cite{Ryu:2006ef}
\be
\label{min_surface}
r^2 + \eta^2 = h^2, 
\ee
where $h$ is a constant, is a solution to the above equations. More concretely, one can explicitly write, 
\be
\cos \chi = \frac{r}{\sqrt{r^2 + \eta^2}}, \quad \sin \chi =  \frac{\eta}{\sqrt{r^2 + \eta^2}}. 
\ee
This we interpret as a positive sign that we have identified a valid calibration. It remains to be seen if this is the only solution. The analysis for the strip is similar and one recovers the expected result \cite{Ryu:2006ef}. We postpone a more in-depth analysis of higher-dimensional examples to future work.

\section*{Acknowledgements} 
We are grateful to Thiago Araujo, Chris Hull, Calin Lazaroiu, Yong-Geun Oh, Chanyong Park and Dieter Van den Bleeken for discussion. We thank Wei Song for comments on the draft. IB is partially supported by the Russian Government programme for the competitive growth of Kazan Federal University. NSD is partially supported by the Scientific and Technological Research Council of Turkey (T\"ubitak) project 116F137. The work of HY is supported in part by National Natural Science Foundation of China, Project 11675244.

\appendix

\section{Geodesics in rotating BTZ} 
\label{sec:geo_BTZ}
Here we solve the geodesic equation for rotating BTZ. It allows us to verify that there is a well-defined boundary interval with spacelike separation, a task that was not fully completed in \cite{Sheikh-Jabbari:2016unm, Colgain:2016lxi} (but see also \cite{Anninos:2013nja}).

As explained in the body of the text, the non-extremal, rotating BTZ metric may be rewritten as (\ref{rotating_BTZ}). In these coordinates, the geodesic equation becomes
\be
\begin{aligned}
0 &= \ddot{X}_- +  \frac{2 r}{r^2 - r_+^2} \dot{X}_- \dot{r}, \\
0 &= \ddot{X}_+ + \frac{2 r}{r^2 - r_-^2} \dot{X}_+ \dot{r}, \\
0 &= \ddot{r} + \frac{(r^2-r_+^2)(r^2-r_-^2)}{r (r_+^2-r_-^2)} \dot{X}_-^2  - \frac{(r^2-r_+^2)(r^2-r_-^2)}{r (r_+^2-r_-^2)} \dot{X}_{+}^2 - \frac{(r^4 - r_+^2 r_-^2) \dot{r}^2}{r (r^2-r_+^2)(r^2-r_-^2)}. 
\end{aligned}
\ee
It is clear from the above equations that it is consistent to truncate so that $X_- = 0$, but we postpone this step until after we have found the general geodesic. To solve the equations, we employ the usual strategy. First, since $X_{\mp}$ are isometry directions, we can integrate the first two equations to identify two constants of motion \footnote{See for example appendix B of \cite{Colgain:2016lxi}.}. Secondly, we can replace the final equation with the requirement that the geodesic be spacelike, 
\be
 - \frac{(r^2-r_+^2)}{(r_+^2 - r_-^2)} \dot{X}_-^2 + \frac{(r^2 - r_-^2)}{r_+^2 - r_-^2} \dot{X}_{+}^2 + \frac{r^2\, \dot{r}^2}{(r^2 - r_+^2)(r^2-r_-^2)}  = 1. 
\ee
Finally, we replace $\dot{X}_{\mp}$ with their conserved quantities and integrate to solve for $r$ in terms of the affine parameter. Suppressing further details, we merely quote the result: 
the solution to the geodesic equation is 
\be
r = \frac{1}{2} \sqrt{ \gamma \cosh 2 s + \alpha}, \quad 
X_{\pm} = \frac{1}{2} \log \left( \frac{e^{2 s} + e^{\Delta_{\pm}}}{e^{2 s} + e^{- \Delta_{\pm}}} \right) + x^{(2)}_{\pm}, 
\ee
where $\Delta_{\pm} = x^{(1)}_{\pm} - x^{(2)}_{\pm}$ and we have further defined, 
\be
\gamma = \frac{8 (r_+^2 - r_-^2) e^{\Delta_+ + \Delta_-}}{(e^{\Delta_+} - e^{\Delta_-})(e^{\Delta_+ + \Delta_-} -1)}, \quad
\alpha = \frac{4 r_+^2 e^{\Delta_-} ( 1+ e^{2 \Delta_+}) - 4 r_-^2 e^{\Delta_{+}} ( 1+ e^{2 \Delta_-})}{(e^{\Delta_+} - e^{\Delta_-})(e^{\Delta_+ + \Delta_-} -1)}. 
\ee

For simplicity, we now set $X_- = 0$ through the choice $x^{(1)}_- = x^{(2)}_- = 0$. Dropping subscripts, the simplified solution then reads: 
\be
\begin{gathered}
r = \frac{1}{2} \sqrt{ \gamma \cosh 2 s + \alpha}, \quad 
X = \frac{1}{2} \log \left( \frac{e^{2 s} + e^{\Delta}}{e^{2 s} + e^{- \Delta}} \right) + x^{(2)}, \\
\gamma = \frac{2 (r_+^2 - r_-^2)}{\sinh^2 (\frac{\Delta}{2} )}, \quad
\alpha = \frac{2( r_+^2 \cosh \Delta - r_-^2)}{\sinh^2 (\frac{\Delta}{2})}. 
\end{gathered}
\ee
Note, as $s \rightarrow \pm \infty$, $X \rightarrow x^{(2)}$ and $X \rightarrow x^{(1)}$, respectively, thus ensuring that our geodesic makes contact with the boundary at two points. This ensures the geodesic is valid from the perspective of holographic entanglement entropy. 

Since $X$ and $r$ are functions of the affine parameter $s$, we can eliminate it to write $X$ directly in terms of $r$ as 
\be
\cosh ( X - \frac{1}{2} (x^{(1)}+x^{(2)}) ) = \cosh \frac{\Delta}{2} \sqrt{\frac{r^2 - r_+^2}{r^2 - r_-^2}}. 
\ee
By shifting $X$ by a constant $X \rightarrow X + \frac{1}{2} (x^{(1)}+x^{(2)})$
the above equation may be simply written as 
\be
\cosh X =  \cosh \frac{\Delta}{2} \sqrt{\frac{r^2 - r_+^2}{r^2 - r_-^2}}. 
\ee

As a consistency check on the result, we note that one can recover equation (B.7) of \cite{Colgain:2016lxi} with $X_- = 0$, provided the independent constant there, namely $L$, is related to $\Delta$ in the following way, 
\be
\label{L_Delta}
L = - r_+ \frac{(e^{\Delta} +1)}{(e^{\Delta} -1)}~~ \Rightarrow~~ \frac{L}{\sqrt{L^2-r_+^2}} =  \cosh \frac{\Delta}{2}, 
\ee 
This shows the result is consistent with earlier analysis. 
 

\end{document}